\documentclass{PoS}

\newcommand{\pipe}{\mbox{\texttt{rPICARD}}}
\newcommand{\repo}{\href{https://bitbucket.org/M_Janssen/Picard}{https://bitbucket.org/M\_Janssen/Picard}}

\title{\pipe{}: A CASA-based Calibration Pipeline for VLBI Data}

\ShortTitle{\pipe{}: A CASA-based Calibration Pipeline for VLBI Data}

\author{\speaker{Michael Janssen}, Ciriaco Goddi, Heino Falcke, Daan van Rossum\\
        Department of Astrophysics/IMAPP\\
        Radboud University, P.O. Box 9010, 6500 GL Nijmegen, The Netherlands.\\
        E-mail: \email{M.Janssen@astro.ru.nl}}

\author{Ilse van Bemmel, Mark Kettenis, Des Small\\
        Joint Institute for VLBI ERIC (JIVE)\\
        Oude Hoogeveensedijk 4, 7991 PD Dwingeloo, The Netherlands.}
        
\author{Ivan~Mart\'i-Vidal\\
        Observatorio de Yebes\\
        IGN, Cerro de la Palera S/N, 19141 Yebes (Guadalajara), Spain.}

\abstract{Currently, HOPS and AIPS are the primary choices for the time-consuming process
of (millimeter) Very Long Baseline Interferometry (VLBI) data calibration.
However, for a full end-to-end pipeline, they either lack the ability to perform 
easily scriptable incremental calibration or do not provide full control over the
workflow with the ability to manipulate and edit calibration solutions directly.
The  Common Astronomy Software Application (CASA) offers all these abilities, together with a secure development future and an
intuitive Python interface, which is very attractive for young radio astronomers.
Inspired by the recent addition of a global fringe-fitter, the capability to convert
FITS-IDI files to measurement sets, and amplitude calibration routines based on
ANTAB metadata, we have developed the the CASA-based
Radboud PIpeline for the Calibration of high Angular Resolution Data (\pipe{}).
The pipeline will be able to handle data from multiple arrays: EHT, GMVA, VLBA
and the EVN in the first release. Polarization and phase-referencing calibration are
supported and a spectral line mode will be added in the future.
The large bandwidths of future radio observatories ask for a scalable reduction
software. Within CASA, a message passing interface (MPI) implementation
is used for parallelization, reducing the total time needed for processing.
The most significant gain is obtained for the time-consuming fringe-fitting task where 
each scan be processed in parallel.}

\FullConference{14th European VLBI Network Symposium \& Users Meeting (EVN 2018)\\
		8-11 October 2018\\
		Granada, Spain}

\begin{document}

\section{Introduction}

The Astronomical Image Processing System (AIPS) \cite{Greisen2003}
has been established as the de-facto standard calibration tool for Very Long Baseline Interferometry (VLBI) data.
However, the AIPS software support is diminishing, while its successor, the Common Astronomy
Software Application (CASA) \cite{McMullin2007} (formally AIPS++), became the main software
for the reduction of connected interferometry data from the Very Large Array (VLA) \cite{Thompson1980}
and the Atacama Large Millimeter/sub-\\millimeter Array (ALMA) \cite{Wootten2009}.
Thanks to an initiative from the \href{https://blackholecam.org/}{BlackHoleCam} \cite{Goddi2017} project
in collaboration with the Joint Institute for VLBI ERIC, CASA has recently received a
VLBI upgrade; new VLBI functionalities like a Schwab-Cotton global fringe-fitter \cite{Schwab1983} and ANTAB-based a-priori calibration routines have been added to the software suite \cite{Bemmel2018}.

The newer CASA software offers some advantages over AIPS, making
it a more attractive choice as a future VLBI data reduction package:
\begin{enumerate}
\item The IPython user interface significantly reduces the learning curve 
for the new `Python' generation of radio astronomers \cite{Momcheva2015}.
\item CASA was designed to support batch processing. The internal data
structure allows full control over all data reduction steps.
\item Parallel processing is supported, allowing the 
computational time to scale with available CPU power
-- an important feature for future observations with
large bandwidths or a wide field of view.
\end{enumerate}

The aforementioned CASA features are ideal prerequisites for a pipeline-based
data reduction.
We have developed a modular, MPI-parallelized, VLBI calibration pipeline based on CASA,
called the Radboud PIpeline for the Calibration of high Angular Resolution Data (\pipe{}).
\pipe{} v1.0.0 is able to handle data in FITS-IDI\footnote{
See \href{https://fits.gsfc.nasa.gov/registry/fitsidi/AIPSMEM114.PDF}{https://fits.gsfc.nasa.gov/registry/fitsidi/AIPSMEM114.PDF}
for a description of the fits-idi data format.}
or Measurement Set (MS)\footnote{See
\href{https://casa.nrao.edu/Memos/229.html}{https://casa.nrao.edu/Memos/229.html}
for a description of the current Measurement Set format.} format.
Simple polarization calibration and phase-referencing are possible in the first release and
spectral line data calibration will be added soon.
The purpose of the pipeline is to provide science-ready visibility data and thereby make VLBI more accessible to non-experts in the community.
Every step of the pipeline is well documented and produces verbose diagnostic
output, such that users should understand how and if the data calibration worked.
It should be noted that a robust open-source VLBI calibration pipeline will bolster
the reproducibility of scientific results.

\pipe{} is hosted on \repo{}.
A specific CASA branch that contains the latest versions of the VLBI calibration functionalities is used. This branch is also open-source; a link to the latest version is given in the README file of the \pipe{} repository.
Eventually, this VLBI branch will be merged into the CASA master branch.

\section{The CASA Calibration Framework}

The CASA calibration framework is a mature and well-tested tool kit, where
all radio interferometric calibration functionalities required to calibrate ALMA and VLA data are in place.
With the recent additions of a fringe-fitter and special amplitude calibration tasks \cite{Bemmel2018}, CASA's
capabilities have been extended to the field of VLBI.

CASA's calibration routines are based on the Hamaker-Bregman-Sault Measurement Equation \cite{Hamaker1996}
to calibrate full Stokes complex visibilities. 
Data corruption effects are corrected with incremental calibration tasks.
The incremental on-the-fly application of previously obtained amplitude calibration solutions
can be used to adjust the weights within the calibration solution intervals.
Measurements with high system temperatures are down-weighted in this process for example.
When available, CASA will make use of per-channel weights.

CASA stores visibilities and metadata as binary tables in the MS data structure.
Raw FITS-IDI visibility data can be converted to a MS with the CASA \textit{importfitsidi} task.

\section{\pipe{} Calibration}

   \begin{figure}
   \centering
   \includegraphics[width=0.85\textwidth]{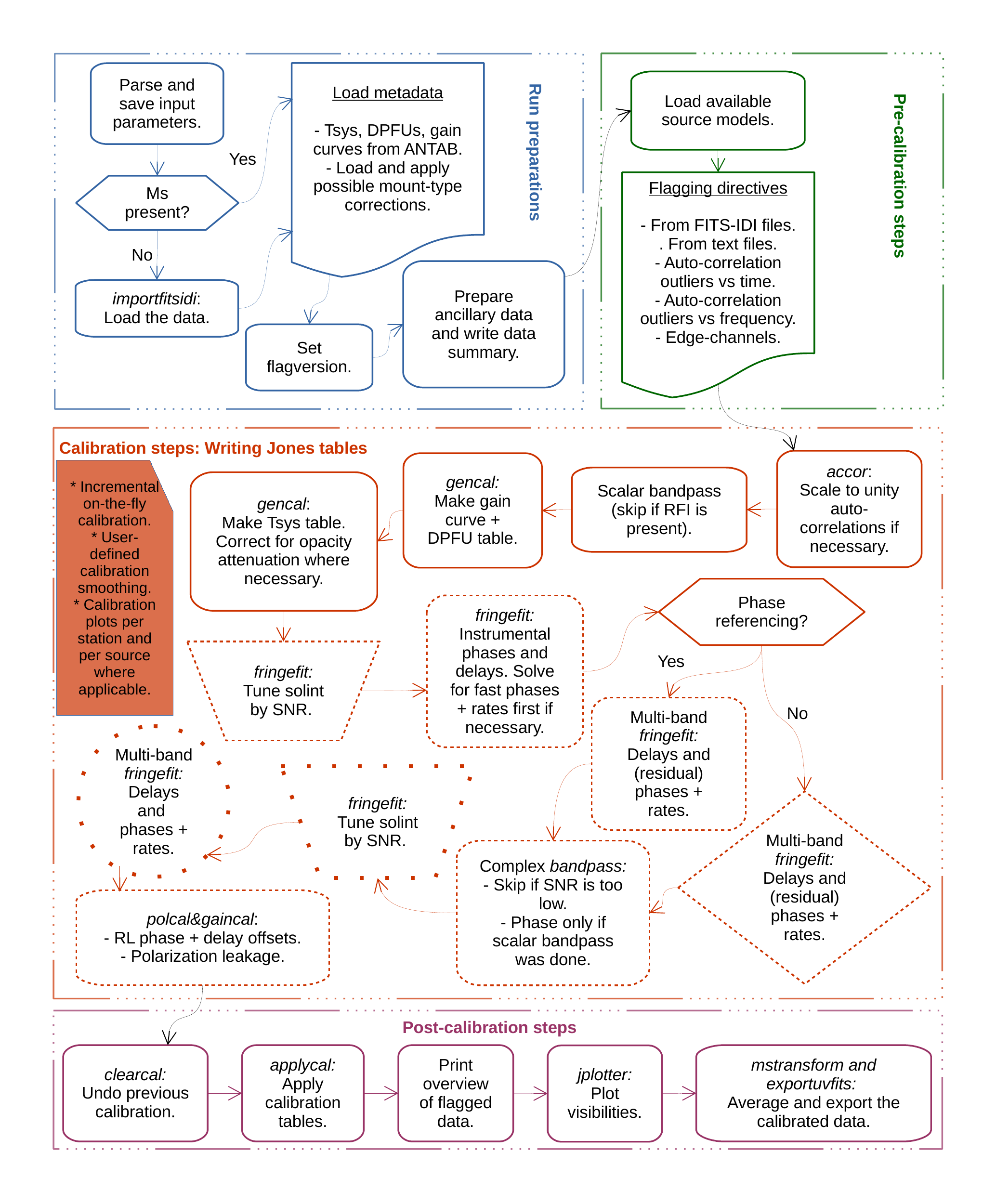}
   \caption{Overview of the \pipe{}'s calibration scheme.
            In the central calibration part, the following correspondences
            are implied based borders of the boxes:
            Solid, dashed, and dotted borders correspond to 
            calibration steps where all sources, only calibrators,
            and only science targets were used to obtain the calibration
            solutions respectively.
            The shape of the boxes indicate to which sources the solutions
            are applied:
            Rectangles -- all sources, diamonds -- calibrators only,
            circles -- science targets only, trapezoids -- solutions are
            not applied to the data.
            }
              \label{calib_flowchart}
    \end{figure}

Figure \ref{calib_flowchart} shows an overview of the \pipe{} calibration flow.
After an initial data flagging, the \textit{accor} task is used to correct for 
erroneous recorder sampler thresholds and the scalar bandpass correction is
performed based on the auto-correlations. This is done with a custom Python
script with the use of basic CASA \textit{tools}.
The amplitude calibration is done based on standard ANTAB metadata, which
is used to compute the time- and frequency-dependent system equivalent flux
densities based on the gains, gain curves, and system temperatures of each station.
The necessary scripts for this a-priori calibration can be found on
\href{https://github.com/jive-vlbi/casa-vlbi}{https://github.com/jive-vlbi/casa-vlbi}.
Where necessary, an additional correction for the signal attenuation from the
atmospheric opacity can be performed by \pipe{}.

The CASA \textit{fringefit} task is used for the phase calibration.
To correct for intra-scan atmospheric effects at short wavelengths, optimal solution intervals
are determined as a compromise between the atmospheric coherence time and the minimum required
integration time to obtain fringe detections. This automated solution interval estimation
is one of the reasons why \pipe{} works for many different arrays without manual parameter
fine-tuning. Additionally, different antennas can be calibrated on different timescales in the
same scan, meaning long integration times can be used for low signal-to-noise ratio (SNR) baselines, while phases distortions on high
SNR baselines can be calibrated on short timescales.
The bright calibrator sources are processed first to solve for instrumental phase and delay offsets
between frequency bands and -- if the SNR is sufficient -- for phase bandpass responses.
After these instrumental effects have been calibrated out, the science targets
are fringe-fitted to obtain a maximum number of detections.

The \pipe{} calibration philosophy is that every calibration step can be fine-tuned, while
the default input parameters will already provide high quality data
for all supported arrays. The pipeline can be used as a very versatile calibration
framework thanks to numerous configuration options, detailed in \cite{Janssen2019}.
\pipe{} produces many diagnostics for each run that can be used to judge the 
quality of the calibration. For example, plots of calibration solutions, lists
of flags that were applied to the data, and log files with detailed information
about every step of the pipeline. Simple command line arguments can be used to
re-run individual steps of the pipeline.
The idea is that users typically calibrate their datasets in a first pass
with the default input parameters. Inspection of all diagnostic output should
reveal the characteristics of the dataset and show which calibration steps did
not produce sufficient solutions. Based on this information the user may
compile a list of flagging statements identifying bad (un-calibratable) data.
A text file containing the flagging commands will be read and applied to the
data automatically by the pipeline.
Next, parameters may be adjusted to improve the calibration
(e.g., fringe search windows or the set of calibrator sources used to correct
for the various instrumental effects).

An advantage of CASA over older calibration packages like AIPS is the built-in
Message Passing Interface (MPI) scalability with CPU cores. In this framework,
\pipe{} will partition the measurement set such that each scan can be processed
in parallel. Especially for the fringe-fitting steps, this yields a significant
speed-up in computational time. The CPU scalability will be crucial for the 
calibration of future experiments with very large bandwidths and wide-field
observations.

\section{\pipe{} Imaging}

Traditional CLEAN imaging of VLBI datasets is often a tedious and interactive process, where
large gain uncertainties and low SNR measurements require multiple rounds of careful
self-calibration iterations, data flagging, and placements of CLEAN windows.

A `semi-automated' imager has been implemented in \pipe{}, independently
of the data calibration framework. The imager is able to image data from
measurement sets or UVFITS files. In essence, it is a wrapper around the 
CASA \textit{tclean} imaging function, performing incremental
imaging and self-calibration loops in an automated way.
With \textit{tclean}, advanced imaging methods such as 
multi-scale multi-frequency synthesis reconstruction methods
and CLEAN auto-boxing are available. The automated
placement of CLEAN boxes was developed for ALMA data.
For VLBI data, the algorithm is able to recover bright and
compact source features, while faint extended emission
will often not be captured. Therefore, manual CLEAN 
boxes placement is still needed in many cases.
The scripted self-calibration loops and automatic
storage of used CLEAN boxes and flags are nonetheless
very valuable for the reproducibility of scientific results.

More details about the \pipe{} calibration and imaging strategies can be found in \cite{Janssen2019}.

\subsection*{Acknowledgements}

This work is supported by the ERC Synergy Grant "BlackHoleCam:  Imaging the Event Horizon of Black Holes" (Grant 610058).

\end{document}